# How are CS students using resources and AI tools for coding tasks?

How do CS students use resources and AI for coding?

Natalia Echeverry

University of Pittsburgh, nae81@pitt.edu

Arun Lekshmi Narayanan,

School of Computing and Information, University of Pittsburgh, arl122@pitt.edu

A survey of 26 CS students reveals that AI coding assistants are mainly used for writing code (second to online searches) while AI chatbots are the top resource for debugging. Participants with different coding experience prefer online help over direct human help from peers and instructors.

## 1  INTRODUCTION

Widespread access to generative AI tools in recent years has brought both excitement and worry to educators in CS courses, given the versatility of such tools to write, edit, explain, and debug code in multiple programming languages. Studies on the use of AI tools in CS courses focus on prescriptive uses of AI tools that preemptively assign a use case to the tool. To our knowledge, this is the first user study that surveys CS students on how they use AI tools for their coding tasks by personal choice.

We surveyed 26 CS students and practitioners with various programming experiences using AI tools for coding tasks (i.e., write, debug, etc.). When asked about the most common resources they have used to write a 300-line program from scratch, blog entries (e.g., Stack Exchange, etc.) were their top choice, followed by AI coding assistants (e.g., GitHub Copilot, etc.). When asked about resources to debug code, AI chatbots (e.g., ChatGPT, etc.) were the most common choice, followed by blog entries.

AI coding assistants are used more for writing code, while AI chatbots are used for debugging tasks. Respondents with all programming experience prefer online resources for coding tasks–whether AI-powered or not– rather than direct human help from peers and instructors.

## 2  AI TOOLS FOR CODING TASKS

Writing and debugging code are core tasks in CS courses. Writing code involves generating a solution (or multiple sub-solutions) that follows the logic and syntax of the programming language. The written program needs to run and produce the desired output. Debugging code includes identifying the portions or concepts in the code that contribute to faulty results.

Before the public release of code-capable AI tools and their widespread availability, online resources —blog entries, reference books, documentation, and human help from instructors and peers— were the go-to for CS students when working on coding tasks. Nowadays, whether instructors approve of them or not, AI tools are a relevant addition to their toolbox.

AI tools are ubiquitous, and access barriers are marginal for most CS students in the United States. Coding assistants integrated into IDEs like GitHub Copilot, Code Whisperer, Tabnine, Ghostwriter, and Codium offer free tier access in educational contexts. Chat LLMs like GPT, Gemini, and LLAMA are free and openly available to anyone with an internet connection.

Schools and instructors have adopted two approaches regarding AI use policies in CS courses. Some have intended to ban it, embrace it, and everything in between [1, 2]. The widespread access to AI tools and their capabilities is difficult to grapple with if there is a good understanding of how this has changed CS students' resource selection. We conducted an online survey and distributed it among CS students in the United States. The survey asked them about the resources they use to do coding tasks.

## 3  ONLINE SURVEY

The questionnaire (A1) was distributed online. It included twelve questions that asked participants about their programming experience, programming languages, resources, and principles they would or currently use for coding tasks. Participants who completed the survey were entered in a raffle for a chance to win a $10 Amazon gift card. We distributed the survey link among a

group of CS students from a research university in the United States and on the Reddit social media platform. The participation criteria were being over 18 years old and having used a programming language in the past six months. Participants who met these criteria could continue answering the survey questions.

The survey remained open for ten days, and we received 26 responses. More than half of respondents were in the 18-24 age range (65.4%), followed by 25-34 (30.8%) and 45-54 (3.8%). Around half of the participants chose Python (53.8%) as the programming language they use most often, followed by Java (34.6%), C++ (7.7%), and Bash (3.8%). One respondent noted that Assembly was not listed in the options and mentioned using Assembly, Bash, and C equally. This suggests future surveys could include Assembly and the "Other" option, allowing respondents to type in additional languages and select more than one programming language. Most respondents said they have more than 3 years (38.5%) of experience using the programming language they selected initially, followed by 1 year (30.8%), 2 years (15.4%), 6 months (7.7%), and less than 6 months (7.7%). After asking about their experience, respondents were asked, 'Can you write a program of about 300 lines of code in the programming language of your choice from scratch?' to which they answered yes (69.2%), no (11.5%), and I haven't tried (19.2%).

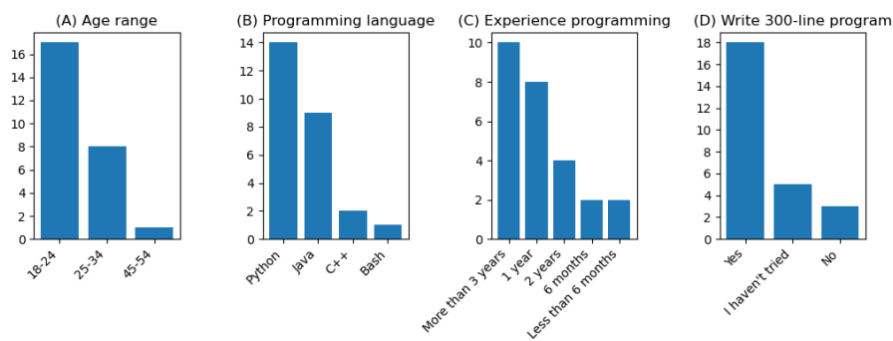

**Figure 1:** (A) Please select your age range. (B) What is the programming language you use most often? (C) How long have you been using this programming language? (D) Can you write a program of about 300 lines of code in the programming language of your choice from scratch?

To the question, 'In which context do you primarily use this programming language? Please select all that apply,' most respondents selected undergraduate coursework (16, 61.5%), followed by personal projects (10, 38.5%), my current job (26.9%), graduate coursework (6, 23.1%), informal learning (19.2%) and other (3.8%) (Figure 2).

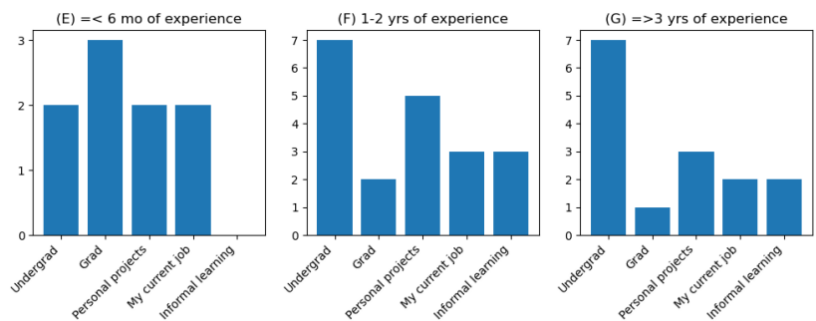

**Figure 2:** In which context do you primarily use this programming language? Please select all that apply.

After completing the programming experience questions, respondents answered questions about their practice. They were asked to type their answer to the question, 'Which code editor or integrated development environment (IDE) would you use to write a 300-line program from scratch?' Respondents reported using IDEs such as Visual Studio Code (11), 'Spyder' (1), 'PyCharm' (2), 'Eclipse' (1), text and source code editors 'Notepad++' (1) and 'Sublime Text' (2). The following two questions in the survey asked about additional resources they use to write and debug a 300-line program. Respondents mostly used online resources such as blog entries (e.g., Stack Exchange), AI coding assistants, and LLMs to help them write a 300-line program from scratch. They also sought help from peers and instructors (Figure 1).

**Table 1: Resources for writing and debugging code (n=26).**

|  | (A) Resources for writing code | (B) Resources for debugging code |
|---|---|---|
|  | Blog entries with a search engine = 19 | AI chatbot=12 |
|  | AI coding assistants=14 | Blog entries=10 |
|  | LLM that generates and discusses code=12 | AI coding assistants=10 |
|  |  | LLM in the cloud or locally = 10 |
|  | Human help=10 | Human help=6 |
|  | Memory=4 | Memory=6 |
|  | Personal files=1 |  |

All the respondents who reported that they can't write a 300-line program from scratch have one year or less than 6 months of experience using the programming language. All of them use or would use Visual Studio Code. The online resources they would use are AI coding assistants, LLMs that generate and discuss code, and blog entries to help them write the program. None reported seeking help from peers and instructors or consulting personal files (Figure 3). Similarly, they said they would use these online resources to debug the program, and this time, all of them included code-capable AI chatbots.

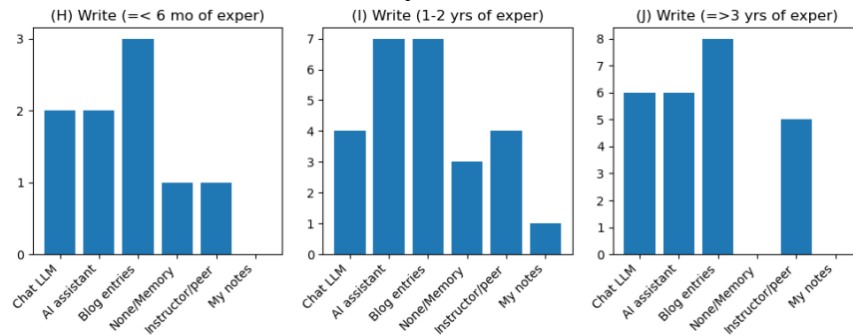

Figure 3: What additional resources would you use to help you write the [300-line] program from scratch? Select all that apply.

Among respondents who have not tried to write a 300-line program from scratch, their experience using the programming language varies. It ranges from 6 months to over 3 years, with a median of 1 year. They all use the programming language (Python=3, Java=2) for their undergraduate work. All except one use Visual Studio Code (VSC=4, Spyder=1). Only one respondent reported seeking help from a colleague, friend, or instructor. Respondents who said they can't or haven't tried to write a 300-line program from scratch use the programming language as part of their undergraduate coursework and prefer online resources rather than human help.

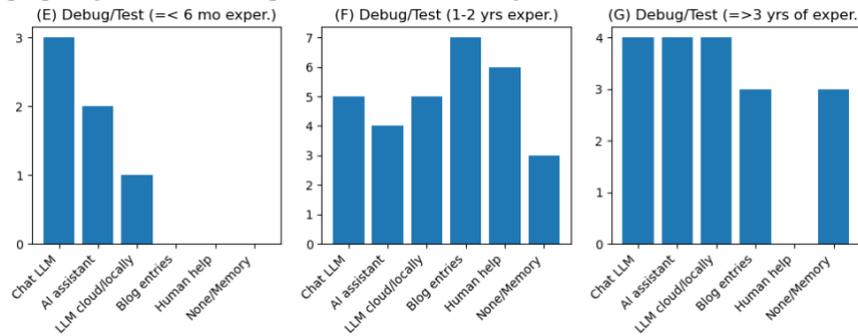

Figure 4: What resources would you most likely use to help you test and debug the 300-line program you wrote from scratch? Select all that apply.

We asked them about the resources they use to work on programming-related tasks. Generative AI chatbots (e.g., ChatGPT, Gemini, Claude AI, Meta AI WhatsApp, etc.) were the most used resources across these tasks: "code comments and explanations" (11), "debug code" (13), and "problem decomposition" (13). AI coding assistants (e.g., GitHub Copilot, Kite, Codeium, IntelliCode.) ranked highly for "code comments and explanations" (11). In contrast, the option "I don't use AI applications for that task." ranked highly for "Problem Decomposition" (11), "Code Quality" (11), and "Coding Style / Patterns" (13). "LLM running in the cloud (e.g., GPT, LLAMA, etc.)" ranked low across all tasks, which may be due to the costs associated with using LLMs online.

## 4 FUTURE WORK

Relying exclusively on studies that assume controlled access to AI tools ignores today's reality. Focusing solely on instructors' views and practices is a one-sided exploration that can miss valuable insight from the students' side. Given the widespread access to AI tools and their ease of use, studying how students use them independently for coding tasks can provide valuable insight to schools, instructors, and AI tool designers.

For future work, we aim to revise the questionnaire (A1) to include more options for programming languages. We also want to translate the questionnaire into multiple natural languages, such as Spanish and French, to expand the survey reach to CS students across the Americas and the Caribbean. This is also partly because of the availability of multilingual LLMs that allow users to use their native language for prompting. The survey results will further guide CS student contextual interviews to explore their resource use, views, and practices.

## 5 CONCLUSIONS

AI coding assistants are used more for writing code, while AI chatbots are used for debugging tasks. Whether AI-powered or not, online resources for coding tasks are preferable to human help from instructors and peers. Exploring how CS students use AI tools for coding tasks can provide valuable insight to schools, instructors, and AI tool designers (especially those in educational technology) so that they can better support them in their learning.

# A APPENDICES

A.1 Questionnaire: Research Study on Tools for Learning and Practicing Computer Programming (2024/11/1)

# Research Study on Tools for Learning and Practicing Computer Programming

The following questions will ask you about the programming languages you use, your experience level, the tools you use, and any formal or informal programming training you have participated in.

This survey will take you less than 5 minutes.

* Indicates required question

1. Please select your age range: *

   *Mark only one oval.*

   ◯ Under 18    *Skip to section 3 (End of Survey)*
   ◯ 18-24
   ◯ 25-34
   ◯ 35-44
   ◯ 45-54
   ◯ 55 or older

2. Have you used any programming languages in the past 6 months? *

   *Mark only one oval.*

   ◯ Yes
   ◯ No    *Skip to section 3 (End of Survey)*

## Programming Experience

3. What is the programming language you use most often? Please select one from the list below.

*Mark only one oval.*

- ○ Python
- ○ C++
- ○ Java
- ○ PHP
- ○ Typescript (Javascript)
- ○ C#
- ○ Bash

4. In which context do you primarily use this programming language? Please select all * that apply.

*Check all that apply.*

- ☐ Undergraduate course work
- ☐ Graduate course work
- ☐ Informal learning
- ☐ My current job
- ☐ Personal projects
- ☐ Other: ______________

5. How long have you been using this programming language? *

*Mark only one oval.*

- ○ Less than 6 months
- ○ 6 months
- ○ 1 year
- ○ 2 years
- ○ More than 3 years

6. Can you write a program of about 300 lines of code in the programming language of your choice from scratch?

   *Mark only one oval.*

   ◯ Yes   *Skip to question 7*

   ◯ No    *Skip to question 7*

   ◯ I haven't tried   *Skip to question 7*

   End of Survey

   I'm sorry, you are not eligible for this study!

## Programming Practice

The questions below ask about the coding principles and tools you would likely use when writing a program from scratch.

7. Which code editor or integrated development environment (IDE) would you use to * write a 300 lines program from scratch? Please type your answer below.

8. What additional resources would you use to help you write the program from scratch? Select all that apply.

   *Check all that apply.*

   ☐ Any large language model (LLM) that I can prompt to generate and discuss code.
   ☐ Any AI coding assistant that will give me autocomplete suggestions as I type.

- [ ] Blog entries on Stack Exchange or other coding sites that I come across after browsing on a search engine.
- [ ] I ask a friend, colleague, or instructor for help.
- [ ] None. I can write a 300-line program from memory.
- [ ] Other: _______________

9. What resources would you most likely use to help you test and debug the 300-line* program you wrote from scratch? Select all that apply.

*Check all that apply.*

- [ ] Any AI chatbot that I can prompt to generate and discuss code.
- [ ] Any AI coding assistant that will give me autocomplete suggestions as I type.
- [ ] Any large language model (LLM) that I can use in the cloud or locally.
- [ ] Blog entries on Stack Exchange or other coding sites that I come across after browsing on a search engine.
- [ ] I ask a friend, colleague or instructor for help.
- [ ] None. I can debug the 300-line program by reviewing it myself.
- [ ] Other: _______________

10. Which tool(s) would you use to help you with the tasks below? Select all that apply.

*Check all that apply.*

| | AI coding assistant (e.g., GitHub Copilot, Kite, Codeium, IntelliCode, etc.) | Generative AI chatbot (e.g., ChatGPT, Gemini, Claude AI, Meta AI WhatsApp, etc.) | LLM running locally (e.g., GPT, LLAMA, etc.) | LLM running in the cloud (e.g., GPT, LLAMA, etc.) | I don't use AI applications for that task. |
|---|---|---|---|---|---|
| Code | | | | | |
| Comments and Explanations | ☐ | ☐ | ☐ | ☐ | ☐ |
| Debug Code | ☐ | ☐ | ☐ | ☐ | ☐ |
| Problem Decomposition | ☐ | ☐ | ☐ | ☐ | ☐ |
| Code Quality | ☐ | ☐ | ☐ | ☐ | ☐ |
| Coding Style / Patterns | ☐ | ☐ | ☐ | ☐ | ☐ |

11. Rate the following coding principles based on their relevance to you, from most relevant to least relevant.

   *Mark only one oval per row.*

| | 0 = Not Important At All | 1 = Of Little Importance | 2 = Of Average Importance | 3 = Very Important | 4 = Absolutely Essential |
|---|---|---|---|---|---|
| KISS (Keep It Simple, Stupid) | ○ | ○ | ○ | ○ | ○ |
| DRY (Don't Repeat Yourself) | ○ | ○ | ○ | ○ | ○ |
| YAGNI (You Aren't Gonna Need It) | ○ | ○ | ○ | ○ | ○ |
| TDA (Tell, Don't Ask) | ○ | ○ | ○ | ○ | ○ |
| SOLID, object oriented programming principles | ○ | ○ | ○ | ○ | ○ |

## Contact Information for the Gift Card Raffle

Please provide your email address and U.S. phone number to participate in the raffle for a $10 Amazon gift card.

12. We would like to invite you to participate in our study about computer programming tools for learning and practice. Would you like to receive more information about the study?

    *Mark only one oval.*

    ◯ Yes

    ◯ No

13. Email Address for the Gift Card Raffle

14. U.S. Phone Number

Thank you for completing this survey!